\documentstyle[12pt,a4,epsfig]{article}

\begin{document}
\newcommand{\be}{\begin{equation}}
\newcommand{\ee}{\end{equation}}
\newcommand{\ba}{\begin{eqnarray}}
\newcommand{\ea}{\end{eqnarray}}
\newcommand{\tr}{\mbox{tr}}
\newcommand{\rauthors}[1]{#1}
\newcommand{\rjournal}[1]{{\it #1}}
\newcommand{\rvolume}[1]{\mbox{\bf #1}}
\newcommand{\ryear}[1]{\mbox{(#1)}}
\newcommand{\rpage}[1]{\mbox{#1}}
\newcommand{\journref}[5]{\rauthors{#1}\ \rjournal{#2}\ \rvolume{#3}\ %
\ryear{#4}\ \rpage{#5}}
\newcommand{\ve}{\varepsilon}
\newcommand{\ch}{\mbox{ch}}
\newcommand{\sh}{\mbox{sh}}
\newcommand{\appex}{ \setcounter{equation}{0}\section}
\newcommand{\figitem}[1]{\item[{\bf Fig.~\ref{#1}}]}
\newcommand{\figcaption}{\refstepcounter{figure}\par\noindent
\centerline{\large\bf Figure \arabic{figure}}\par}

7/Dec/99\hfill FIAN/TD-25/99\\

\vspace{1cm}

\begin{center}
{\Large\bf NLO correction to one-particle inclusive production at
high energies.}
\end{center}
\medskip
\begin{center}
{\bf Dmitry Ostrovsky\footnote{e-mail: ostrov@td.lpi.ac.ru}}\\
\medskip
{\it P.N.~Lebedev Physical Institute, 117924 Leninsky pr. 53, Moscow,
Russia}
\end{center}
\bigskip
\begin{center}
{\bf Abstract}\\
Next-to-leading order correction to the one-particle inclusive
cross section in the framework of
 high energy factorization is calculated. Numerical results for
midrapidity region are compared with predictions of conventional
calculations based on collinear factorization.

\end{center}
\medskip

\newpage

\section{Introduction}

The recently completed calculation of the NLO correction to BFKL Pomeron
\cite{NLOBFKL1} involves as its ingredients the formulae
for the cross section of tree level two-particle production and
one-loop virtual correction to one-particle production in Quasi Multi Regge
Kinematics (QMRK).
These results can be used in studying  the properties of the
inclusive two-particle production at leading order \cite{BG,ELR+}
and of the one-particle inclusive cross section in the
next-to-leading order at high energies.

The total cross section in the next-to-leading order(NLO BFKL) was calculated
in \cite{NLOBFKL1} and is now under an extensive discussion
\cite{NLOBFKL2,CCS}.  The total cross section has a complicated
structure due to delicate cancellations of singularities in various virtual
contributions  and infrared divergencies arising in integrations over
parameters of real correction.  Cancellation between divergent parts of real
and virtual contributions is also needed for computation of the next-to-leading
order one-particle inclusive production cross section, but here the structure
of this cancellation is simpler and more transparent than in the case  of total
cross section.

The problem of one-particle inclusive production in high energy
hadron collision was investigated, in the leading order (LO),
in a number of works  \cite{ELR+}.
The purpose of this paper is to calculate  one-particle inclusive
cross section to the next-to-leading order at high energies
proceeding as far as possible with analytical calculations and then
turning to numerical estimates.

The outline of the paper is as follows.

In the second section we briefly review the results on the particle production
cross sections obtained within the high energy factorization scheme and
compare them to the expressions obtained using collinear factorization.

  In the third section we describe a calculation leading to the explicit
expression for the one-particle inclusive production cross section in the
next-to-leading order.

In Section 4 some numerical results on particle production in
central rapidity region are presented and discussed.

Section 5 contains a brief conclusion.

\section{Particle production at high energies}\label{factorization}

From the theoretical viewpoint the hadron scattering at very high energies is
special in the way the hard degrees of freedom (partons) are formed from
colliding hadrons. When the ratio of hardness of the process $k_\perp$
(which is the transverse momentum of produced particle)
to the invariant energy of colliding hadrons, $\sqrt{S}$,
is not too small (up to $10^{-2}$) it is possible
to describe the structure functions of hadrons by taking into account only
processes that contribute logarithms of $k_\perp/\Lambda_{QCD}$ at leading
order, i.e. by resummation of the $\alpha_s^n\ln^n(k_\perp/\Lambda_{QCD})$
terms.  Such structure function is given DGLAP evolution equation \cite{DGLAP}.

Since the emergence  of the combination $\alpha_s\,\ln(k_\perp/\Lambda_{QCD})$
implies strong ordering of emitted particles
in their transverse momenta up until a hard collision block, the
transverse momentum of detected particle $k_\perp$ is parametrically
bigger  than that of any parton involved in the process.
Therefore it is possible to calculate the cross section of the hard process
using  the initial on-shell partons. This prescription is known as collinear
factorization \cite{CollFact} and leads to the well-known result for the
production rate:
\be\label{CFLO}
\frac{d\sigma}{dk^2dy_1}=
2\int dy_2 x_1f_a(x_1,k^2)\frac{d\hat{\sigma}_{ab}}{dk^2}x_2f_b(x_2,k^2),
\ee
where $f_a(x,k^2)$ is a structure function for the parton of type
$a$ and $x_{1,2}=k_\perp (e^{\pm y_1}+e^{\pm y_2})/\sqrt{S}$.

At high energies for particles produced with $k_\perp\ll\sqrt{S}$
another big logarithm, $\ln(1/x)$, is important.
The resummation of such logarithmic contributions can
become more important than of $\ln(k^2/\Lambda^2_{QCD})$.
The resummation of the leading energy logarithms for the structure function is
described by BFKL equation  \cite{BFKL}.  The domain of validity of the BFKL
equation in describing the structure functions is at present not well
understood.
It is likely that for some kinematical region the correct approach
is to resum the logarithms of both types, or at least interpolate between
two types of resummation as done, e.g., in the  CCFM\cite{CCFM} equation.

The main point is that at high energies the transverse momenta of the incoming
parton fluxes can no longer be neglected. To take them into account a new
approach called $k_\perp$ or high energy factorization was proposed
 \cite{TransFact, CCH}.  Extensive description of the method and  various
applications can be found in  \cite{CCH}.  Let us note  that this method was
{\it de facto} used earlier in  \cite{LR}.

The method of high energy factorization is based on consideration of
"partons" with nonzero transverse momentum that are, in contrast with the
traditional collinear factorization case, virtual particles.
In this case colliding hadrons are described by
unintegrated structure function,
$\phi$, so that
\be\label{phi}
\phi(x,q^2)=q^2\frac{\partial xg(x,q^2)}{\partial q^2},
\ee
where $\phi/q^2$ is proportional to the probability to
find the incident parton (essentially a gluon, because gluons give leading
contribution at high energies)
with the longitudinal momentum component $xp_a$ ($p_a$ is a momentum
of the incident  particle) and transverse momentum component $q_\perp $.

Note that such an interpretation is somewhat  oversimplified, because
due to the quantum structure of QCD evolution some form-factors
may arise changing the form of $\phi$. For DGLAP evolution it is a
Sudakov form-factor \cite{DDT}.
However, when studying the
semi-inclusive quantity like one-particle production cross section it is
legitimate
to use the unintegrated structure function in the simple form of
Eq.(\ref{phi})\footnote{I am grateful to Yu.L.Dokshitzer
for pointing me this out}.

Scattering of the off-shell "partons" are described by generalized
cross sections calculated in Quasi Multi Regge Kinematics (QMRK)
\cite{FL1,FL}.

In QMRK one studies the  $2\rightarrow n+2$ scattering process
with two outgoing particles having  almost the same momenta as the incident
ones and remaining $n$ particles emitted into the central rapidity region
separated by large rapidity gaps from incoming particles (the situation, where
rapidity gaps between $n$ particles are also large, corresponds to
Multi Regge Kinematics, MRK).
Large rapidity gaps allow to distinguish  the quantities related
to the incident particles from  those describing the cross section
of the hard process of particle production in the central rapidity region.

Let us for example consider the cross section of the process
$gg\rightarrow ggg$  in the limit of high energy:

\be\label{2g3g}
\frac{d\sigma_{gg\rightarrow ggg}}{d^2k_\perp dy}=
\frac{4N^3_c\alpha^3_s}{\pi^2 (N^2_c-1)}
\int\frac{d^2q_{1\perp}}{q^2_{1\perp}}
\frac{\delta^{(2)}(q_{1\perp}+q_{2\perp}-k_{\perp})}{k^2_\perp}
\frac{d^2q_{2\perp}}{q^2_{2\perp}},
\ee
where $q_{1,2}=p_{a,b}-p^\prime_{a,b}$. in Eq.~(\ref{2g3g})
the above-mentioned factorization of the cross section is clearly seen. Indeed,
the first, second and third factors under the integral  correspond to
$p_a\rightarrow p^\prime_a,q_1$ splitting, $q_1, q_2\rightarrow k$
"scattering" and $p_b\rightarrow p^\prime_2,q_2$ splitting respectively.

The factors related to the splitting of the incident particles
should further be transformed to structure functions. This can be done in
two steps. First, one assembles  incident partons into wave
packets describing by form factors \cite{BG}.
The second step is taking into account additional radiation and
virtual corrections summed to give the unintegrated structure
functions $\varphi(x,q_\perp)$ with $x$ fixed  by the kinematics
of the considered process.

The cross sections of producing $n=1,2$ particles
in the central region to the lowest perturbative order read:

$$
\frac{d\sigma_1}{d^2k_\perp dy}=
          \int d^2q_{1\perp} d^2q_{2\perp}
               \frac{\varphi(x_{1,0},q_{1\perp})}{q^2_{1\perp}}
               \frac{d\hat{\sigma}_1}{dk^2_\perp}
               \frac{\varphi(x_{2,0},q_{2\perp})}{q^2_{2\perp}},
$$
\be\label{sigma1}
\frac{d\hat{\sigma}_1}{dk^2_\perp}=\frac{4N_c\alpha_s}{N_c^2-1}
\frac{\delta^{(2)}(q_{1\perp}+q_{2\perp}-k_{\perp})}{k^2_{\perp}},
\ee
$$
      x_{1,0}=k_\perp e^{y}/\sqrt{S},\quad x_{2,0}=k_\perp e^{-y}/\sqrt{S};
$$

$$
\frac{d\sigma_2}{d^2k_{1\perp} d^2k_{2\perp}dy_1 dy_2}=
        \int d^2q_{1\perp} d^2q_{2\perp}
             \frac{\varphi(x_{1},q_{1\perp})}{q^2_{1\perp}}
             \frac{d\hat{\sigma}_2}{d^2k_{1\perp} d^2k_{2\perp} d\Delta y}
             \frac{\varphi(x_{2},q_{2\perp})}{q^2_{2\perp}},
$$
\be\label{sigma2}
\frac{d\hat{\sigma}_2}{d^2k_{1\perp} d^2k_{2\perp} d\Delta y}=
\frac{2N_c^2\alpha_s^2}{(N_c^2-1)\pi^2}
\frac{\delta^{(2)}(q_{1\perp}+q_{2\perp}-k_{1\perp}-k_{2\perp})}
{q^2_{1\perp} q^2_{2\perp}} {\cal A},
\ee
$$
x_1=k_{1\perp}e^{y_1} (1 + k_{2\perp}e^{ \Delta y})/\sqrt{S},\;
x_2=k_{1\perp}e^{-y_1}(1 + k_{2\perp}e^{-\Delta y})/\sqrt{S}.
$$
$$
\Delta y=y_2-y_1.
$$

A part of the analytical expression for $\cal{A}$  can be found in \cite{FL,FKL}
for the subprocesses $gg\rightarrow gg$ and in \cite{FL,FFFK} for the
$gg\rightarrow q\bar{q}$ ones. The explicit form of
$\cal{A}$ was recently derived in \cite{asymm} and is given in Appendix A.
Note that the formula for $gg\rightarrow q\bar{q}$ cross section
from  \cite{FFFK} coincides
with the analogous formula in  \cite{CCH} in the limit of massless quarks.

Eq.(\ref{sigma1}) gives the rate of one-particle production in the leading
order. It was studied in a number of publications  \cite{ELR+}.
Our aim is to calculate first correction to it.

\section{Real and virtual contributions to NLO one-particle production}

The one-particle production in the next-to-leading order includes two
contributions, real and virtual. The real contribution comes from the
two-particle cross section Eq.(\ref{sigma2}) integrated over the phase
space of one of the particles (considered unobservable) and the fixed
four-momentum of the second particle:
\be\label{real_naive}
\frac{d\sigma_r}{d^2k_{1\perp}dy_1}=
        2\int d\Phi
             \frac{\varphi(x_{1},q_{1\perp})}{q^2_{1\perp}}
             \frac{d\hat{\sigma}_2}{d^2k_{1\perp} d^2k_{2\perp} d\Delta y}
             \frac{\varphi(x_{2},q_{2\perp})}{q^2_{2\perp}}
\ee
where
\be
d\Phi=d^2q_{1\perp} d^2q_{2\perp} d^2k_{2\perp} d\Delta y
\ee
The factor of 2 in Eq.~(\ref{real_naive}) reflects the identity of
outgoing particles.

The virtual contribution has the same
form as in Eq.(\ref{sigma1}), but instead of $\hat{\sigma}_1$
we must use
\be\label{virthat}
\frac{d\hat{\sigma}_v}{dk^2_\perp}=\frac{4N_c\alpha_s}{N_c^2-1}
\frac{\delta^{(2)}(q_{1\perp}+q_{2\perp}-k_{\perp})}{k^2_{\perp}}
{\cal V}(q_{1\perp}, q_{2\perp})
\ee
The virtual correction contains  both ultraviolet and infrared divergencies.
The ultraviolet one leads, through standard renormalization procedure,
to the running coupling constant. The infrared divergence must cancel with
the infrared and collinear divergencies in the real contribution.

\subsection{Cancellation of collinear and infrared divergencies}

Let us now  outline at the formal level how this cancellation occurs.
To deal with the collinear singularity we must introduce a
jet defining algorithm which in two particle production case
could be expressed through the function $S(k, k_1, k_2)$
($k_1$, $k_2$ and $k$ are on-shell 4-vectors) so that Eq.(\ref{real_naive})
is replaced by
\be\label{real_S}
\frac{d\sigma_r}{d^2k_{\perp}dy}=
\int d^2k_{1\perp} d^2k_{2\perp} dy_1 dy_2
\frac{d\sigma_2}{d^2k_{1\perp} d^2k_{2\perp}dy_1 dy_2}
S(k, k_1, k_2).
\ee
To provide the sought for cancellation between the real
and virtual corrections $S$ should be an infrared safe quantity,
which means that the following property should hold (cf. \cite{KS}):
\be\label{infra_safe}
S(k,\lambda k_1, (1-\lambda)k_1)=\delta^{(2)}(k_\perp-k_{1\perp})
\delta(y-y_1), \quad 0<\lambda<1
\ee
In the following we choose $S$ in the following form:
\ba\label{jet_def}
S(k, k_1, k_2)&=&\theta(R>R_0)\sum\limits_{i=1,2}
\delta^{(2)}(k_\perp-k_{i\perp})\delta(y-y_i)\\
&+&
\theta(R<R_0)\delta^{(2)}(k_\perp-k_{1\perp}-k_{2\perp})
\delta\left(y-\frac{1}{2}
\ln\frac{k_{1\perp}e^{y_1}+k_{2\perp}e^{y_2}}%
{k_{1\perp}e^{-y_1}+k_{2\perp}e^{-y_2}}\right),\nonumber
\ea
where $R^2=(\phi_1-\phi_2)^2+(y_1-y_2)^2$.
Although the last line in Eq.(\ref{jet_def}) may look artificial, it has
the natural meaning.  If we claim that "combined" particle, formed by two
particles indistinguishable under given resolution,  has definite
rapidity and that 4-momenta of two particles are added up to form
the 4-momentum of "combined" particle, we immediately arrive at
Eq.(\ref{jet_def}). It is straightforward to check that
Eq.(\ref{jet_def}) satisfies Eq.~(\ref{infra_safe}).

Grouping together Eqs.(\ref{sigma2}),(\ref{real_S}) and (\ref{jet_def})
we find:
\ba\label{real}
\frac{d\sigma_r}{d^2k_{\perp}dy}&=&
        2\int d\Phi
             \frac{\varphi(x_{1},q_{1\perp})}{q^2_{1\perp}}
             \frac{d\hat{\sigma}_2}{d^2k_{\perp} d^2k_{2\perp} d\Delta y}
             \frac{\varphi(x_{2},q_{2\perp})}{q^2_{2\perp}}
        \theta(R>R_0)\nonumber\\
&+&     \int d\Phi
             \frac{\varphi(\tilde{x}_{1},q_{1\perp})}{q^2_{1\perp}}
             \frac{d\hat{\sigma}_2}{d^2k_{1\perp} d^2k_{2\perp} d\Delta y}
             \frac{\varphi(\tilde{x}_{2},q_{2\perp})}{q^2_{2\perp}}
        \theta(R<R_0),
\ea
where $k_{1\perp}=k_{\perp}-k_{2\perp}$ and
$\tilde{x}_{1,2}=e^{\pm y}\sqrt{\Sigma/S}$ with
$\Sigma=k^2_1+k^2_2+2k_1k_2\ch(\Delta y)$ (see Appendix A),
and $k_i=|k_{i\perp}|$ which is a notation we shall use from now on.
Note that when $R\rightarrow 0$ $\tilde{x}_{1}\rightarrow x_{1,0}$
and $\tilde{x}_{2}\rightarrow x_{2,0}$. Let us now rewrite the second
term in Eq.(\ref{real}) as:
$$
        \left[\int d\Phi
             \left(
             \frac{\varphi(\tilde{x}_{1},q_{1\perp})}{q^2_{1\perp}}
             \frac{\varphi(\tilde{x}_{2},q_{2\perp})}{q^2_{2\perp}}-
             \frac{\varphi({x}_{1,0},q_{1\perp})}{q^2_{1\perp}}
             \frac{\varphi({x}_{2,0},q_{2\perp})}{q^2_{2\perp}}
             \right)\right.
             \frac{d\hat{\sigma}_2}{d^2k_{1\perp} d^2k_{2\perp} d\Delta y}
$$
\be\label{temp}
      +\left.
        \int d\Phi
             \frac{\varphi({x}_{1,0},q_{1\perp})}{q^2_{1\perp}}
             \frac{d\hat{\sigma}_2}{d^2k_{1\perp} d^2k_{2\perp} d\Delta y}
             \frac{\varphi({x}_{2,0},q_{2\perp})}{q^2_{2\perp}}
\right]
        \theta(R<R_0).
\ee
The integration in the first line is free from divergencies, while
in the second line integrals over $d^2k_{2\perp}$ and $d\Delta y$
do not involve structure functions and could be done analytically.
Before doing this integration we make a replacement
$\theta(R<R_0)=1-\theta(R>R_0)$ in the last term and then substitute
Eq.(\ref{temp}) to Eq.(\ref{real}) to yield:
\ba\label{real_2}
&&\frac{d\sigma_r}{d^2k_{\perp}dy}=\nonumber\\
&&      \int d^2q_{1\perp} d^2q_{2\perp}
             \frac{\varphi({x}_{1,0},q_{1\perp})}{q^2_{1\perp}}
             \frac{\varphi({x}_{2,0},q_{2\perp})}{q^2_{2\perp}}
             \int d^2k_{2\perp} d\Delta y
             \frac{d\hat{\sigma}_2}{d^2k_{1\perp} d^2k_{2\perp} d\Delta y}
\nonumber\\
&&
       +\int d\Phi
         \left(
           2 \frac{\varphi(x_{1},q_{1\perp})}{q^2_{1\perp}}
             \frac{d\hat{\sigma}_2}{d^2k_{\perp} d^2k_{2\perp} d\Delta y}
             \frac{\varphi(x_{2},q_{2\perp})}{q^2_{2\perp}}
             \theta(R(k,k_2)>R_0)\right.\\
&&\qquad\quad -\left.
             \frac{\varphi(x_{1,0},q_{1\perp})}{q^2_{1\perp}}
             \frac{d\hat{\sigma}_2}{d^2k_{1\perp} d^2k_{2\perp} d\Delta y}
             \frac{\varphi(x_{2,0},q_{2\perp})}{q^2_{2\perp}}
             \theta(R(k_1,k_2)>R_0)
         \right)
         \nonumber \\
&&     +\int d\Phi
             \left(
             \frac{\varphi(\tilde{x}_{1},q_{1\perp})}{q^2_{1\perp}}
             \frac{\varphi(\tilde{x}_{2},q_{2\perp})}{q^2_{2\perp}}-
             \frac{\varphi({x}_{1,0},q_{1\perp})}{q^2_{1\perp}}
             \frac{\varphi({x}_{2,0},q_{2\perp})}{q^2_{2\perp}}
             \right)\nonumber\\
&&\qquad\qquad \frac{d\hat{\sigma}_2}{d^2k_{1\perp} d^2k_{2\perp} d\Delta y}
\theta(R<R_0)\nonumber,
\ea
where we indicate that
$R(k,k_2)=((y-y_2)^2+(\phi-\phi_2)^2)^{1/2}$ and
$R(k_1,k_2)=((y_1-y_2)^2+(\phi_1-\phi_2)^2)^{1/2}$ are different in
the third and forth lines of Eq.(\ref{real_2}).
The third line in Eq.(\ref{real_2}) has an
infrared singularity when $k_2\rightarrow 0$ and the forth one has
singularities when $k_2\rightarrow 0$ or $k_1\rightarrow 0$.
However, it is possible to combine the singularities in the forth line
so that we find only one singular
point and also an accompanying factor of two thus providing  a
cancellation of singularities in the third and forth lines.

Indeed, the expression under the integral
in the forth line in Eq.(\ref{real_2}) is symmetric under the
simultaneous transformation
$k_{1\perp}\leftrightarrow k_{2\perp}$ and
$\Delta y\leftrightarrow -\Delta y$ which is nothing else than the
permutation of the two produced particles.
Therefore we can simply multiply this term at
$2\theta(k_{1\perp}>k_{2\perp})$ to obtain:
\ba\label{real_fin}
&&\frac{d\sigma_r}{d^2k_{\perp}dy}=\nonumber\\
&&      \int d^2q_{1\perp} d^2q_{2\perp}
             \frac{\varphi({x}_{1,0},q_{1\perp})}{q^2_{1\perp}}
             \frac{\varphi({x}_{2,0},q_{2\perp})}{q^2_{2\perp}}
             \int d^2k_{2\perp} d\Delta y
             \frac{d\hat{\sigma}_2}{d^2k_{1\perp} d^2k_{2\perp} d\Delta y}
\nonumber\\
&&
       +2\int d\Phi
         \left(
             \frac{\varphi(x_{1},q_{1\perp})}{q^2_{1\perp}}
             \frac{d\hat{\sigma}_2}{d^2k_{\perp} d^2k_{2\perp} d\Delta y}
             \frac{\varphi(x_{2},q_{2\perp})}{q^2_{2\perp}}
             \theta(R(k,k_2)>R_0)\right.\\
&&\quad -\left.
             \frac{\varphi(x_{1,0},q_{1\perp})}{q^2_{1\perp}}
             \frac{d\hat{\sigma}_2}{d^2k_{1\perp} d^2k_{2\perp} d\Delta y}
             \frac{\varphi(x_{2,0},q_{2\perp})}{q^2_{2\perp}}
             \theta(R(k_1,k_2)>R_0)\theta(k_{1\perp}>k_{2\perp})
         \right)
         \nonumber \\
&&     +\int d\Phi
             \left(
             \frac{\varphi(\tilde{x}_{1},q_{1\perp})}{q^2_{1\perp}}
             \frac{\varphi(\tilde{x}_{2},q_{2\perp})}{q^2_{2\perp}}-
             \frac{\varphi({x}_{1,0},q_{1\perp})}{q^2_{1\perp}}
             \frac{\varphi({x}_{2,0},q_{2\perp})}{q^2_{2\perp}}
             \right)\nonumber\\
&&\qquad\qquad \frac{d\hat{\sigma}_2}{d^2k_{1\perp} d^2k_{2\perp} d\Delta y}
\theta(R<R_0).\nonumber
\ea

The combination of the third and forth lines in Eq.(\ref{real_fin})
is free from singularities (note that for $k_2\rightarrow 0 $ there is no
difference between $R(k,k_2)$ and $R(k_1,k_2)$) and the singularity in
the second line cancels with that in ${\cal V}$ in Eq.(\ref{virthat}).
Combining Eq.(\ref{real_fin}) with Eqs.(\ref{sigma1}),(\ref{virthat})
we obtain the second order correction to the one-particle inclusive
production in the high energy factorization scheme:
\ba\label{pre-main}
&&\frac{d\sigma^{(2)}}{d^2k_{\perp}dy}=\nonumber\\
&&      \int d^2q_{1\perp} d^2q_{2\perp}
             \frac{\varphi({x}_{1,0},q_{1\perp})}{q^2_{1\perp}}
             \frac{\varphi({x}_{2,0},q_{2\perp})}{q^2_{2\perp}}
             \left(
             \frac{d\hat{\sigma}_2}{d^2k_{\perp}}
             +\frac{d\hat{\sigma}_v}{d^2k_\perp}\right)
\nonumber\\
&&
       +2\int d\Phi
         \left(
             \frac{\varphi(x_{1},q_{1\perp})}{q^2_{1\perp}}
             \frac{d\hat{\sigma}_2}{d^2k_{\perp} d^2k_{2\perp} d\Delta y}
             \frac{\varphi(x_{2},q_{2\perp})}{q^2_{2\perp}}
             \theta(R(k,k_2)>R_0)\right.\\
&&     -\left.
             \frac{\varphi(x_{1,0},q_{1\perp})}{q^2_{1\perp}}
             \frac{d\hat{\sigma}_2}{d^2k_{1\perp} d^2k_{2\perp} d\Delta y}
             \frac{\varphi(x_{2,0},q_{2\perp})}{q^2_{2\perp}}
             \theta(R(k_1,k_2)>R_0)\theta(k_{1\perp}>k_{2\perp})
         \right)
         \nonumber \\
&&     +\int d\Phi
             \left(
             \frac{\varphi(\tilde{x}_{1},q_{1\perp})}{q^2_{1\perp}}
             \frac{\varphi(\tilde{x}_{2},q_{2\perp})}{q^2_{2\perp}}-
             \frac{\varphi({x}_{1,0},q_{1\perp})}{q^2_{1\perp}}
             \frac{\varphi({x}_{2,0},q_{2\perp})}{q^2_{2\perp}}
             \right)\nonumber\\
&&\qquad\qquad \frac{d\hat{\sigma}_2}{d^2k_{1\perp} d^2k_{2\perp} d\Delta y}
\theta(R<R_0).\nonumber
\ea

Let us now analyse the MRK limit of Eq.(\ref{pre-main}), that is we take limit
$\Delta y\rightarrow \infty $ in ${\cal A}$ (see Eq.(\ref{sigma2}) and
Eq.(\ref{gggg}) in Appendix). In this limit
\be
{\cal A}\rightarrow {\cal A}_{MRK}=
\frac{q^2_{1\perp}q^2_{2\perp}}{k^2_{1\perp}k^2_{2\perp}}
\ee
which is precisely the combination of two leading order BFKL kernels
responsible for real particles production. However, let us remind that
the leading order one-particle production in high-energy factorization,
described by Eq.(\ref{sigma1}) includes MRK contributions
to all orders if the unintegrated
structure function, $\phi(x,q_\perp)$ includes resummation to
all orders of $\alpha_s\ln(1/x)$. It is evidently the case for
structure functions undergoing BFKL equation. For other types
of structure functions we just assume  that the
resummed $\alpha^n_s\ln^n(1/x)$ terms are included in some hidden way.
Consequently, we must subtract
${\cal A}_{MRK}$ from ${\cal A}$, and we will imply this subtraction
in the following.

To proceed further we must calculate
\be\label{difficult}
\frac{d\hat{\sigma}_2}{d^2k_\perp}= \int d^2k_{2\perp} d\Delta y
\frac{d\hat{\sigma}_2}{d^2k_{1\perp} d^2k_{2\perp} d\Delta y}.
\ee

Note that the quantity in brackets in the first line of Eq.(\ref{pre-main})
should coincide (up to the constant factors depending on normalization)
with the NLO BFKL kernel written explicitly in  \cite{NLOBFKL1}.
Note however, that when calculating the real contribution  in  \cite{FKL} the
terms vanishing after integration over $d^2k_\perp$ were dropped which did not
change the result for NLO BFKL Pomeron itself. In calculating the one-particle
inclusive cross sections these contributions have to be kept.

The integration in Eq.(\ref{difficult}) is very difficult.
Fortunately, we can do it not
for whole $\hat{\sigma}_2$, but only for its singular part
$\hat{\sigma}^s_2$. We also have to change some other terms in
Eq.(\ref{pre-main}) that emerge when arriving from Eq.(\ref{real})
at Eq.(\ref{pre-main}). Finally, the result is
\ba\label{main}
&&\frac{d\sigma^{(2)}}{d^2k_{\perp}dy}=\nonumber\\
&&      \int d^2q_{1\perp} d^2q_{2\perp}
             \frac{\varphi({x}_{1,0},q_{1\perp})}{q^2_{1\perp}}
             \frac{\varphi({x}_{2,0},q_{2\perp})}{q^2_{2\perp}}
             \left(
             \frac{d\hat{\sigma}^s_2}{d^2k_{\perp}}
             +\frac{d\hat{\sigma}_v}{d^2k_\perp}\right)
\nonumber\\
&&
       +2\int d\Phi
         \left(
             \frac{\varphi(x_{1},q_{1\perp})}{q^2_{1\perp}}
             \frac{d\hat{\sigma}_2}{d^2k_{\perp} d^2k_{2\perp} d\Delta y}
             \frac{\varphi(x_{2},q_{2\perp})}{q^2_{2\perp}}
             \theta(R(k,k_2)>R_0)\right.\\
&&     -\left.
             \frac{\varphi(x_{1,0},q_{1\perp})}{q^2_{1\perp}}
             \frac{d\hat{\sigma}^s_2}{d^2k_{1\perp} d^2k_{2\perp} d\Delta y}
             \frac{\varphi(x_{2,0},q_{2\perp})}{q^2_{2\perp}}
             \theta(R(k_1,k_2)>R_0)\theta(k_{1\perp}>k_{2\perp})
         \right)
         \nonumber \\
&&     +\int d\Phi
             \left(
             \frac{\varphi(\tilde{x}_{1},q_{1\perp})}{q^2_{1\perp}}
             \frac{d\hat{\sigma}_2}{d^2k_{1\perp} d^2k_{2\perp} d\Delta y}
             \frac{\varphi(\tilde{x}_{2},q_{2\perp})}{q^2_{2\perp}}
             \right.\nonumber\\
&&\qquad\qquad -   \left.
             \frac{\varphi({x}_{1,0},q_{1\perp})}{q^2_{1\perp}}
             \frac{d\hat{\sigma}^s_2}{d^2k_{1\perp} d^2k_{2\perp} d\Delta y}
             \frac{\varphi({x}_{2,0},q_{2\perp})}{q^2_{2\perp}}
             \right)\theta(R<R_0).\nonumber
\ea

\subsection{Integration.}

Let us now  choose the singular part of
${\cal A}$ with contributions from quark and gluon production added up
in the form (see Appendix, MRK part as it was mentioned above
is subtracted):
\be\label{Asing}
{\cal A}^{s}=-\frac{q_1^2q_2^2}{2k^2_{1}k^2_{2}}
+\frac{q_1^2 q_2^2\ch(\Delta y)}{k_{1}k_{2} s}
-\left(1-\frac{n_f}{4N_c}\right)\frac{2q_1^2 q_2^2}{s\Sigma}
+\left(\frac{D-2}{2}-\frac{n_f}{N_c}\right)\frac{E^2}{8s^2}
\ee
This form of ${\cal A}^{s}$
is chosen to avoid artificial ultraviolet divergency which occurs if
one takes $\Sigma=k^2$ as it is in collinear and infrared limits.
For the same purpose we take $E$ in the form:
\be
E=\frac{1}{\Sigma}[k^2(\bar{q}_1-\bar{q}_2)(\bar{k}_1-\bar{k}_2)
-(q_1^2-q_2^2)(k_1^2-k_2^2)+2k_1k_2\sh(\Delta y)(k^2-q_1^2-q_2^2)]
\ee

Now we integrate the singular part of ${\cal A}$ over transverse
two-dimensional momentum space analytically continued to $D-2=2+2\varepsilon$
or, strictly speaking, with
\be
d^2k_{2\perp}\rightarrow
\frac{d^{2+2\varepsilon}k_{2\perp}}{(2\pi)^{2\varepsilon}}
\ee
and over rapidity.
The results of the calculation of the integral over ${\cal A}^s$ are
given in Appendix B.
The answer reads (see Eq.(\ref{sigma2}) for relation between
$\hat{\sigma}_2$ and $\cal A$)
\ba\label{int_r}
\frac{d\hat{\sigma}^s_r}{d^2k_{\perp}}&=&
\frac{2N_c^2\alpha_s^2}{N_c^2-1}
\frac{\delta^{(2)}(q_{1\perp}+q_{2\perp}-k_{\perp})}{k^2}
\frac{\Gamma(1-\ve)}{(4\pi)^{1+\ve}}
\frac{4\Gamma^2(1+\ve)}{\ve \Gamma(1+2\ve)}
\left(\frac{k^2}{\mu^2}\right)^\ve\\
&&\left(\frac{1}{\ve}+2\psi(1)-2\psi(1+2\ve)-
\frac{11+8\ve}{2(1+2\ve)(3+2\ve)}+
\frac{n_f}{4N_c}\frac{4+6\ve}{(1+2\ve)(3+2\ve)}\right)\nonumber
\ea

The result for $\hat{\sigma}_v$ was derived in  \cite{FLvirt}. Note that
the answer depends on the arrangement of different corrections to
QMRK amplitude. In this paper the symmetric variant  \cite{NLOBFKL1}
is chosen.
\ba\label{int_v}
&&\frac{d\hat{\sigma}_v}{d^2k_{\perp}}=
\frac{4N^2_c\alpha^2_s}{N_c^2-1}
\frac{\delta^{(2)}(q_{1\perp}+q_{2\perp}-k_{\perp})}{k^2}
\frac{\Gamma(1-\ve)}{(4\pi)^{1+\ve}}\times \nonumber\\
&&
\left[-\frac{2}{\ve^2}\left(\frac{k^2}{\mu^2}\right)^\ve
      +\frac{1}{\ve}\left(\frac{11}{3}-\frac{2}{3}\frac{n_f}{N_c}\right)
      +\pi^2
      +\frac{k^2}{3}
      \left\{
              \left(11-2\frac{n_f}{N_c}\right)
              \frac{\ln q^2_1/q^2_2}{q^2_1-q^2_2}+
       \right.
\right. \\
&&
\left.
       \left.
            \left(1-\frac{n_f}{N_c}\right)
            \left(\left(\frac{q^2_1}{q^2_2}-\frac{q^2_2}{q^2_1}
                        -2\ln\frac{q^2_1}{q^2_2}
                  \right)
                  \frac{2q^2_1 q^2_2-\bar{q}_1\bar{q}_2%
                   (q^2_1+q^2_2+4\bar{q}_1\bar{q}_2)}{(q^2_1-q^2_2)^3}
                   +\frac{\bar{q}_1\bar{q}_2}{q^2_1 q^2_2}
            \right)
       \right\}
\right]\nonumber
\ea

From Eqs.(\ref{int_r}) and (\ref{int_v}) it is easy to see that divergencies
of real and virtual parts cancel leaving a finite contribution to
the first line in Eq.(\ref{main}):
\ba
&&\frac{d\hat{\sigma}^s_r}{d^2k_{\perp}}+\frac{d\hat{\sigma}_v}{d^2k_{\perp}}=
\frac{N^2_c\alpha^2_s}{(N_c^2-1)\pi}
\frac{\delta^{(2)}(q_{1\perp}+q_{2\perp}-k_{\perp})}{k^2}
\left[-\left(\frac{11}{3}-\frac{2}{3}\frac{n_f}{N_c}\right)\ln\frac{k^2}{\mu^2}
\right.
\nonumber\\
&&
      -\frac{2\pi^2}{3}+\frac{64}{9}-\frac{7}{9}\frac{n_f}{N_c}
      +\frac{k^2}{3}
      \left\{
              \left(11-2\frac{n_f}{N_c}\right)
              \frac{\ln q^2_1/q^2_2}{q^2_1-q^2_2}+
       \right.
\label{int_rv}\\
&&
\left.
       \left.
            \left(1-\frac{n_f}{N_c}\right)
            \left(\left(\frac{q^2_1}{q^2_2}-\frac{q^2_2}{q^2_1}
                        -2\ln\frac{q^2_1}{q^2_2}
                  \right)
                  \frac{2q^2_1 q^2_2-\bar{q}_1\bar{q}_2%
                   (q^2_1+q^2_2+4\bar{q}_1\bar{q}_2)}{(q^2_1-q^2_2)^3}
                   +\frac{\bar{q}_1\bar{q}_2}{q^2_1 q^2_2}
            \right)
       \right\}
\right]\nonumber
\ea
The first term in Eq.(\ref{int_rv}), which is proportional to $\ln(k^2/\mu^2)$,
is nothing but the well-known contribution corresponding to the running coupling.
Therefore, after replacing $\alpha_s$ by the running coupling $\alpha_s(k^2)$,
one should drop this term.

The Eqs. (\ref{main}) and (\ref{int_rv}) together with Eq.(\ref{Asing}) and
the formula from Appendix A provide an analytical expression for the
one-particle inclusive cross section.

For practical applications one should integrate
over the parameters of unintegrated structure
functions (disregarding trivial elimination of delta-functions).
The corresponding numerical calculations numerical studies will be
descried in the next Section.

\section{Numerical estimates}

The numerical results strongly depend on the type of structure functions
used in the calculation. Let us first consider  the asymptotic BFKL
structure function \cite{fBFKL}
\be\label{fBFKL}
\phi(x,q^2)=C\left(\frac{x_0}{x}\right)^\lambda\frac{q}{q_0}
\frac{1}{\sqrt{\pi\lambda''\ln(x_0/x)}}
\exp\left[-\frac{\ln^2(q^2/q^2_0)}{4\lambda''\ln(x_0/x)}\right],
\ee
with $\lambda=4\ln 2N_c\alpha_s/\pi$ and
$\lambda''=14\zeta(3)N_c\alpha_s/\pi$;
$\alpha_s$ is chosen to be equal $0.2$. Because the asymptotic
BFKL structure function is a solution of the linear homogeneous
equation, it does not have definite normalization and, moreover, the parameters
$q_0$ and $x_0$ are arbitrary. Since the calculation with this structure
function is illustrative only, we simply choose $C=1$, $q_0=1$GeV and $x_0=1$.
From Eq.(\ref{main}) it is clear that the NLO correction to the production
process depends on the parameter $R$ describing the collinear angle.
For this calculation we take $R=0.7$. Furhter discussion of the $R$ dependence
will be given below.

It is important to note that although in asymptotic BFKL structure
function the strong coupling constant does not run, to be consistent we
should however make it run in a semihard vertices (cf. the discussion after
Eq.(\ref{main})). In the actual calculation with 1-loop $\alpha_s$ we choose
$\Lambda_{QCD}=200\mbox{MeV}$.

The one-particle inclusive cross section for $\sqrt{S}=5.5\mbox{TeV}$,
$y=0$, and $n_f=4$ is shown in Fig.~1 where
for collinear factorization Eq.(\ref{CFLO}) was used  with
$$
xg(x,k^2)=\int\limits_0^{k^2}\frac{dq^2}{q^2}\phi(x,q^2)
$$

As our second example we study the production process using
the AKMS structure function  \cite{AKMS}.
In AKMS approach unintegrated structure function is
obtained by application of BFKL equation to the evolution
of structure function on $x$ from $x=10^{-2}$ where $\phi(10^{-2}, k^2)$
is chosen to give the best fit for experimental data.

This structure function can be fitted with an acceptable accuracy
at least in the range of interest $x=10^{-2}..10^{-4}$ by
\be\label{fAKMS}
\phi(x,q^2)=\frac{A}{x^\beta}\frac{q}{q_0}
\exp\left[-\frac{B(x)\ln^2(q^2/q^{\prime 2}(x))}{4\ln(1/x)}\right],
\ee
with parameters $A=8.55\cdot 10^{-2}$, $\beta=0.486$,
$q^\prime=0.758+5.41x^{0.6816}$ and
$B=3.51-2.91\ln\ln(1/x)+0.793(\ln\ln(1/x))^2$.
The results of calculation are given in Fig.~2 again with 1-loop $\alpha_s$,
$\Lambda_{QCD}=200\mbox{MeV}$, $\sqrt{S}=5.5\mbox{TeV}$, $n_f=4$,
$y=0$, and $R=0.7$.

Finally, in Fig.~3 we show the one-particle inclusive cross sections
calculated with GRV94(NLO) structure function \cite{GRV94}
satisfying the DGLAP equation not related to BFKL.
However, it possibly includes leading MRK
contribution through the initial conditions for DGLAP evolution.
Note the rapid increase and broadening of structure function
with decreasing $x$, which is the characteristic property of BFKL induced
structure functions.
This calculation was done for
$\sqrt{S}=1.8, 5.5\, \mbox{and}\, 14\mbox{TeV}$,
2-loop $\alpha_s$ with $\Lambda_{QCD}=200\mbox{MeV}$ (because with
these parameters GRV94(NLO) is calculated), $n_f=4$, $y=0$, and $R=0.7$.

From  Figs.~1,2,3
we see that  taking into account the NLO corrections
leads to the  decrease of the
particle production rate at high energies.
For the structure functions obtained in
BFKL approach NLO corrections change cross sections substantially
(up to $50\%$ in chosen kinematical interval), for non-BFKL
GRV structure function changes are more dramatic: corrected cross
sections are 2 to 5 times smaller than the leading order ones
(apparently this results are sensitive to the cone size).
We have no explanation to this fact. We hope that further studies of
relationships between different types of structure functions may make
this  subject clearer.

Figs.~3a,b,c show that ratios of differential cross sections calculated
to the LO and NLO accuracy in high energy factorization and to the LO in
collinear factorization are insensitive to the c.m.s. energy in the domain
$x\ll 1$.

Finally, in Fig.~4 we show the dependence of NLO cross section on the
cone size $R$. It may be fitted well by function of the
type $A+B\ln{R}+CR$ and becomes infinitely large (and negative)
at $R\rightarrow 0$. This is the general property of quantities with
cancelling virtual and real corrections showing that at small values of $R$
the fixed order perturbation theory is not valid (cf. \cite{EKS}).

\section{Conclusion}

The paper is devoted to the calculation of next-to-leading order correction to
one particle inclusive production in the framework of high-energy
factorization. High energy factorization scheme allows one to account for the
initial transverse momentum of the colliding partons. The natural setup for
particle production processes leading to high energy factorization is provided
by Quasi-Multy Regge Kinematics.

The results of computation of NLO contributions to BFKL Pomeron
(cf.\cite{NLOBFKL1} and references therein)  can be used to compute the
next-to-leading order corrections to one particle inclusive production at high
energies. This correction includes real and virtual pieces.  The infrared
singularity in the virtual piece in the NLO contribution cancels the infrared
singularity in its real one  when an infrared safe jet algorithm is applied.
The explicit calculations of the infrared stable one-particle inclusive cross
section at the next-to-leading order constitutes the main  result of the paper.

Numerical estimates were made to analyze the magnitude of NLO corrections for
typical semihard transverse momenta and central rapidity region. Here we
observe an essential dependence  on the type of the structure functions used in
the computation and shows more stability for BFKL-type structure functions and
than  for DGLAP one.

\begin{center}

\it Acknowledgements
\end{center}
I am grateful to A.V. Leonidov for suggesting the idea of the paper
and also for stimulating and helpful discussions.
Special thanks to O.V. Ivanov for pointing me out the powerful
method of multidimensional integration \cite{Korobov}.

The work was supported by INTAS
within the research program of ICFPM, grant 96-0457.


\setcounter{section}{0}
\renewcommand{\thesection}{{\it Appendix} \Alph{section}}
\renewcommand{\theequation}{\Alph{section}.\arabic{equation}}
\renewcommand{\thesubsection}{\Alph{section}.\arabic{subsection}}

\appex{Cross sections of pair production in high energy factorization}

We will use the following notation:
$$
s=2(k_1k_2\ch(\Delta y)-k_{1\perp}k_{2\perp});
$$
$$
t=-(q_{1\perp}-k_{1\perp})^2-k_1k_2e^{\Delta y},\;
u=-(q_{1\perp}-k_{2\perp})^2-k_1k_2e^{-\Delta y};
$$
$$
\Sigma=x_1x_2S=k_1^2+k_2^2+2k_1k_2\ch(\Delta y),
$$
with $k_1=\sqrt{k_{1\perp}^2}$, $k_2=\sqrt{k_{2\perp}^2}$
and $k_{1\perp}k_{2\perp}$ is the dot product with
2d Euclidean metric.

Combined gluons and quarks (fermions) contribution to $gg$
scattering has the form (adopted to (\ref{sigma2}))
\be\label{Atotal}
{\cal A}={\cal A}_{gluons}+\frac{n_f}{4N^3_c}{\cal A}_{fermions}
\ee

\subsection{$gg\rightarrow gg$}
$$
{\cal A}_{gluons}={\cal A}_1+{\cal A}_2
$$

\begin{eqnarray}
{\cal A}_1=
&q_1^2q_2^2&
    \left\{ -\frac{1}{tu}+\frac{1}{4tu}\frac{q_1^2q_2^2}{k_1^2k_2^2}-
            \frac{e^{\Delta y}}{4tk_1k_2}-\frac{e^{-\Delta y}}{4uk_1k_2}+
            \frac{1}{4k_1^2k_2^2}+
    \right.\nonumber\\
&&         \frac{1}{\Sigma}
             \left[-\frac{2}{s}
                \left(1+k_1k_2(\frac{1}{t}-\frac{1}{u})\sh(\Delta y)
                \right)+
                \frac{1}{2k_1k_2}(1+\frac{\Sigma}{s})\ch(\Delta y)-
             \right.\nonumber\\
&&
               -\frac{q_1^2}{4s}
                      [(1+\frac{k_2}{k_1}e^{-\Delta y})\frac{1}{t}+
                       (1+\frac{k_1}{k_2}e^{\Delta y})\frac{1}{u}]\nonumber\\
&&           \left.
    \left.
               -\frac{q_2^2}{4s}
                      [(1+\frac{k_1}{k_2}e^{-\Delta y})\frac{1}{t}+
                       (1+\frac{k_2}{k_1}e^{\Delta y})\frac{1}{u}]
             \right]
    \right\}
\label{gggg}
\end{eqnarray}
\begin{eqnarray}
{\cal A}_2=\frac{D-2}{4}\left\{\left(
\frac{(k_{1\perp}-q_{1\perp})^2
(k_{2\perp}-q_{1\perp})^2-k_1^2k_2^2}{tu}\right)^2-
\right.\nonumber\\
-\frac{1}{4}
\left.\left( \frac{(k_{2\perp}-q_{1\perp})^2-k_1k_2e^{-\Delta y}}
                  {(k_{2\perp}-q_{1\perp})^2+k_1k_2e^{-\Delta y}}
             -\frac{E}{s}
      \right)
      \left( \frac{(k_{1\perp}-q_{1\perp})^2-k_1k_2e^{\Delta y}}
                  {(k_{1\perp}-q_{1\perp})^2+k_1k_2e^{\Delta y}}
             +\frac{E}{s}
      \right)
\right\},\nonumber
\end{eqnarray}

$$
E=(q_{1\perp}-q_{2\perp})(k_{1\perp}-k_{2\perp})-\frac{1}{\Sigma}
(q_1^2-q_2^2)(k_1^2-k_2^2)+2k_1k_2\sh(\Delta y)
\left(1-\frac{q_1^2+q_2^2}{\Sigma}\right).
$$

\subsection{$gg\rightarrow q\bar{q}$}
$$
{\cal A}_{fermions}=N_c^2{\cal A}_{1f}+{\cal A}_{2f}
$$

\begin{eqnarray}
{\cal A}_{1f}&=&\left\{ 2\frac{q_{1}^{2}q_{2}^{2}}{s\Sigma }
\left(1+k_{1}k_{2}\sh (\Delta y)(\frac{1}{t}-\frac{1}{u})\right)
-\left( \frac{(k_{1\perp}-q_{1\perp})^{2}(k_{2\perp}-q_{1\perp})^{2}
-k_{1}^{2}k_{2}^{2}}{tu}\right)^{2}+\right.   \nonumber \\
&&\left. \frac{1}{2}
      \left( \frac{(k_{2\perp}-q_{1\perp})^2-k_1k_2e^{-\Delta y}}
                  {(k_{2\perp}-q_{1\perp})^2+k_1k_2e^{-\Delta y}}
             -\frac{E}{s}
      \right)
      \left( \frac{(k_{1\perp}-q_{1\perp})^2-k_1k_2e^{\Delta y}}
                  {(k_{1\perp}-q_{1\perp})^2+k_1k_2e^{\Delta y}}
             +\frac{E}{s}
      \right)
\right\}\label{ggqq}
\end{eqnarray}
and
$$
{\cal A}_{2f}=\left\{ \left( \frac{(k_{1\perp}-q_{1\perp})^{2}
(k_{2\perp}-q_{1\perp})^{2}-k_{1}^{2}k_{2}^{2}}{tu}\right)^{2}
-\frac{q_{1}^{2}q_{2}^{2}}{tu}\right\}
$$
where $E$ is the same as for gluons.

\appex{Integrals}

To make integration easily it is worth to
change integration over $\Delta y$ to
integration over $x=k_1/(k_1+k_2e^{\Delta y})$.
\be
\int\limits_{-\infty}^{\infty}d\Delta y\,\dots = \int\limits_0^1
\frac{dx}{x(1-x)}\,\dots
\ee

\ba
\int\frac{d^{2+2\ve}k_{2\perp}}{(2\pi)^{2\ve}}
\left(-\frac{1}{2}\frac{1}{k^2_1k^2_2}\right)&=&
-\frac{1}{2}
\int\frac{d^{2+2\ve}k_{2\perp}}{(2\pi)^{2\ve}}
\frac{1}{(k_\perp-k_{2\perp})^2k^2_{2\perp}}\nonumber\\
&=&
-\pi\frac{\Gamma(1-\ve)}{(4\pi)^\ve}\frac{k^{2\ve}}{k^2}
\frac{\Gamma^2(1+\ve)}{\ve\Gamma(1+2\ve)}
\ea

\ba
\int\frac{d^{2+2\ve}k_{2\perp}}{(2\pi)^{2\ve}}
\frac{\ch(\Delta y)}{k_1k_2s}&=&
\int\frac{d^{2+2\ve}k_{2\perp}}{(2\pi)^{2\ve}}
\left(\frac{1-x}{2k_2^2sx}+\frac{x}{2k_1^2s(1-x)}\right)\\
&=&\int\frac{d^{2+2\ve}k_{2\perp}}{(2\pi)^{2\ve}}
\frac{(1-x)^2}{2k^2_2((1-x)k_\perp-k_{2\perp})^2}+ (x\leftrightarrow 1-x)
\nonumber\\
&=&
\pi\frac{\Gamma(1-\ve)}{(4\pi)^\ve}\frac{k^{2\ve}}{k^2}
\frac{\Gamma^2(1+\ve)}{\ve\Gamma(1+2\ve)}[(1-x)^{2\ve}+x^{2\ve}]
\nonumber
\ea

Let's now combine these two contributions and perform an integration over
$x$
\be
I_{12}=\pi\frac{\Gamma(1-\ve)}{(4\pi)^\ve}\frac{k^{2\ve}}{k^2}
\frac{\Gamma^2(1+\ve)}{\ve\Gamma(1+2\ve)}
\int\frac{dx}{x(1-x)}[(1-x)^{2\ve}+x^{2\ve}-1].
\ee
In order to avoid divergencies we introduce infinitesimal parameter
$\delta$ ($\delta\ll\ve$) so that
\ba
\int\frac{dx}{x(1-x)}[(1-x)^{2\ve}+x^{2\ve}-1]&=&
\lim\limits_{\delta\rightarrow 0}
\int\frac{dx}{x^{1-\delta}(1-x)^{1-\delta}}[(1-x)^{2\ve}+x^{2\ve}-1]
\nonumber\\
=\lim\limits_{\delta\rightarrow 0}
\left(2\frac{\Gamma(\delta)\Gamma(2\ve+\delta)}{\Gamma(2\ve+2\delta)}-
\frac{\Gamma^2(\delta)}{\Gamma(2\delta)}\right)&=&
\frac{1}{\ve}+2\psi(1)-2\psi(1+2\ve)
\ea
and
\be
I_{12}=\pi\frac{\Gamma(1-\ve)}{(4\pi)^\ve}\frac{k^{2\ve}}{k^2}
\frac{\Gamma^2(1+\ve)}{\ve\Gamma(1+2\ve)}
(\frac{1}{\ve}+2\psi(1)-2\psi(1+2\ve))
\ee

For the integrations involving $\Sigma$ it is useful to change
$k_{2\perp}$ on $\kappa=k_{2\perp}-(1-x)k_\perp$ so that
$s=\kappa^2/x(1-x)$ and $\Sigma=\kappa^2/x(1-x)+k^2$.
Now
\be
\int\frac{dx}{x(1-x)}\int\frac{d^{2+2\ve}k_{2\perp}}{(2\pi)^{2\ve}}
\frac{1}{s\Sigma}=
\pi\frac{\Gamma(1-\ve)}{(4\pi)^\ve}\frac{k^{2\ve}}{k^2}
\frac{\Gamma^2(1+\ve)}{\ve\Gamma(1+2\ve)}\frac{1}{1+2\ve}
\ee
and
\be
\int\frac{dx}{x(1-x)}\int\frac{d^{2+2\ve}k_{2\perp}}{(2\pi)^{2\ve}}
\frac{E^2}{8q_1^2q_2^2s^2}=
\pi\frac{\Gamma(1-\ve)}{(4\pi)^\ve}\frac{k^{2\ve}}{k^2}
\frac{\Gamma^2(1+\ve)}{\ve\Gamma(1+2\ve)}\frac{1-\ve}{2(1+2\ve)(3+2\ve)}
\ee

\newpage

\begin{center}
\large\bf Figure captions
\end{center}

All cross sections $d\sigma/d^2kdy$ are calculated for combined gluon and
quark contributions with $n_f=4$, $y=0$

\begin{itemize}

\item[\bf Figure 1] $d\sigma/d^2kdy$ calculated with asymptotic BFKL
structure function Eq.(\ref{fBFKL}).
$\sqrt{S}=5.5\mbox{TeV}$, 1-loop $\alpha_s$ with
$\Lambda_{QCD}=200\mbox{MeV}$, $R=0.7$

\item[\bf Figure 2] $d\sigma/d^2kdy$ calculated with AKMS
structure function Eq.(\ref{fAKMS}).
$\sqrt{S}=5.5\mbox{TeV}$, 1-loop $\alpha_s$ with
$\Lambda_{QCD}=200\mbox{MeV}$, $R=0.7$

\item[\bf Figure 3] $d\sigma/d^2kdy$ calculated with GRV94(NLO)
structure function \cite{GRV94}. 2-loop $\alpha_s$ with
$\Lambda_{QCD}=200\mbox{MeV}$, $R=0.7$\\
{\bf a.} $\sqrt{S}=14\mbox{TeV}$ \\
{\bf b.} $\sqrt{S}=5.5\mbox{TeV}$\\
{\bf c.} $\sqrt{S}=1.8\mbox{TeV}$

\item[\bf Figure 4] $R$-dependence of $d\sigma/d^2kdy$
calculated with GRV94(NLO).
$\sqrt{S}=5.5\mbox{TeV}$, 2-loop $\alpha_s$ with
$\Lambda_{QCD}=200\mbox{MeV}$.

\end{itemize}

\newpage

\textwidth=17cm
\oddsidemargin=0mm
\topmargin=-10mm

\begin{figure}
\begin{center}
\vspace*{0cm}
\epsfig{file=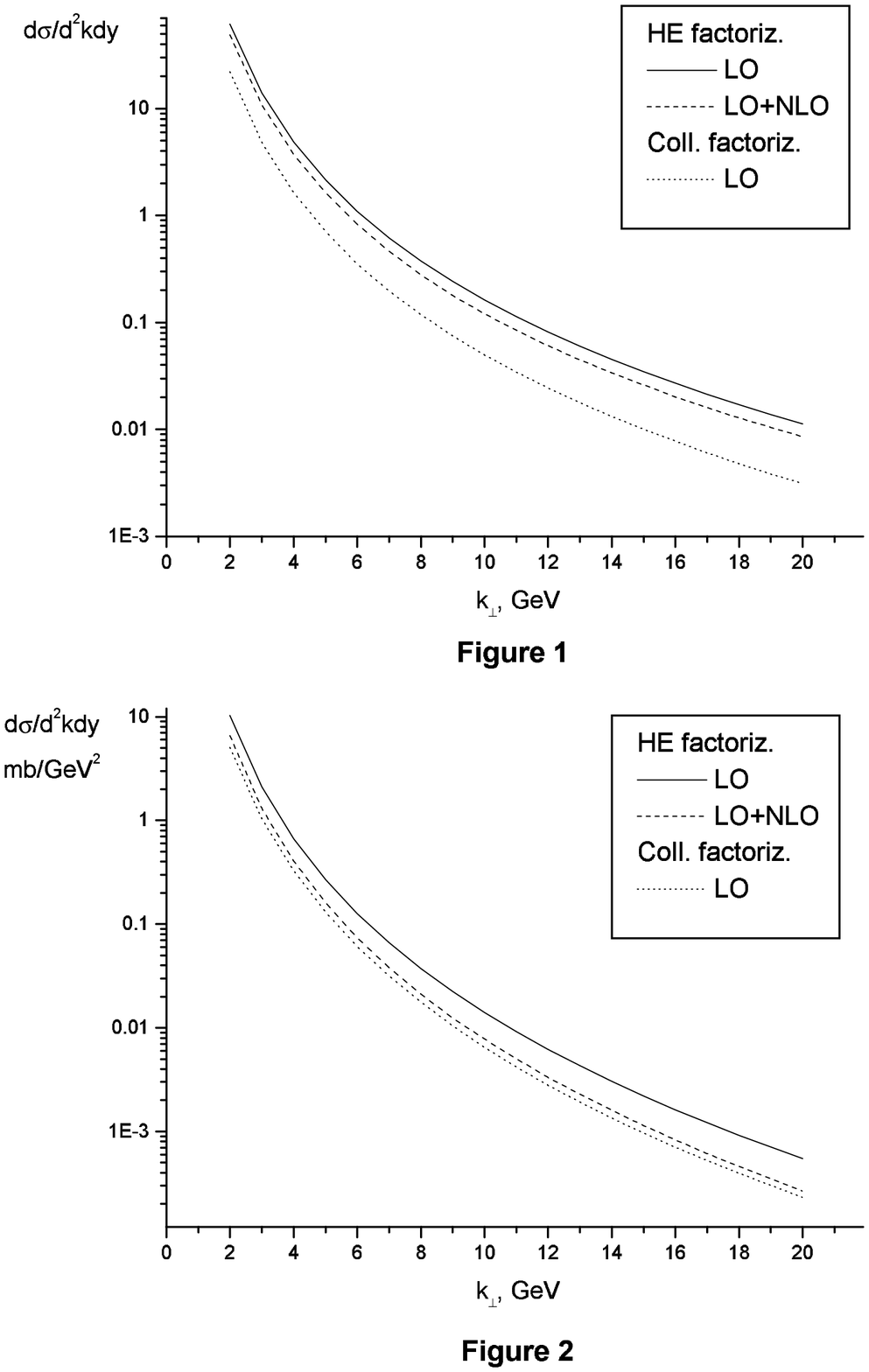,width=12cm}
\end{center}
\end{figure}

\newpage

\begin{figure}
\begin{center}
\vspace*{0cm}
\epsfig{file=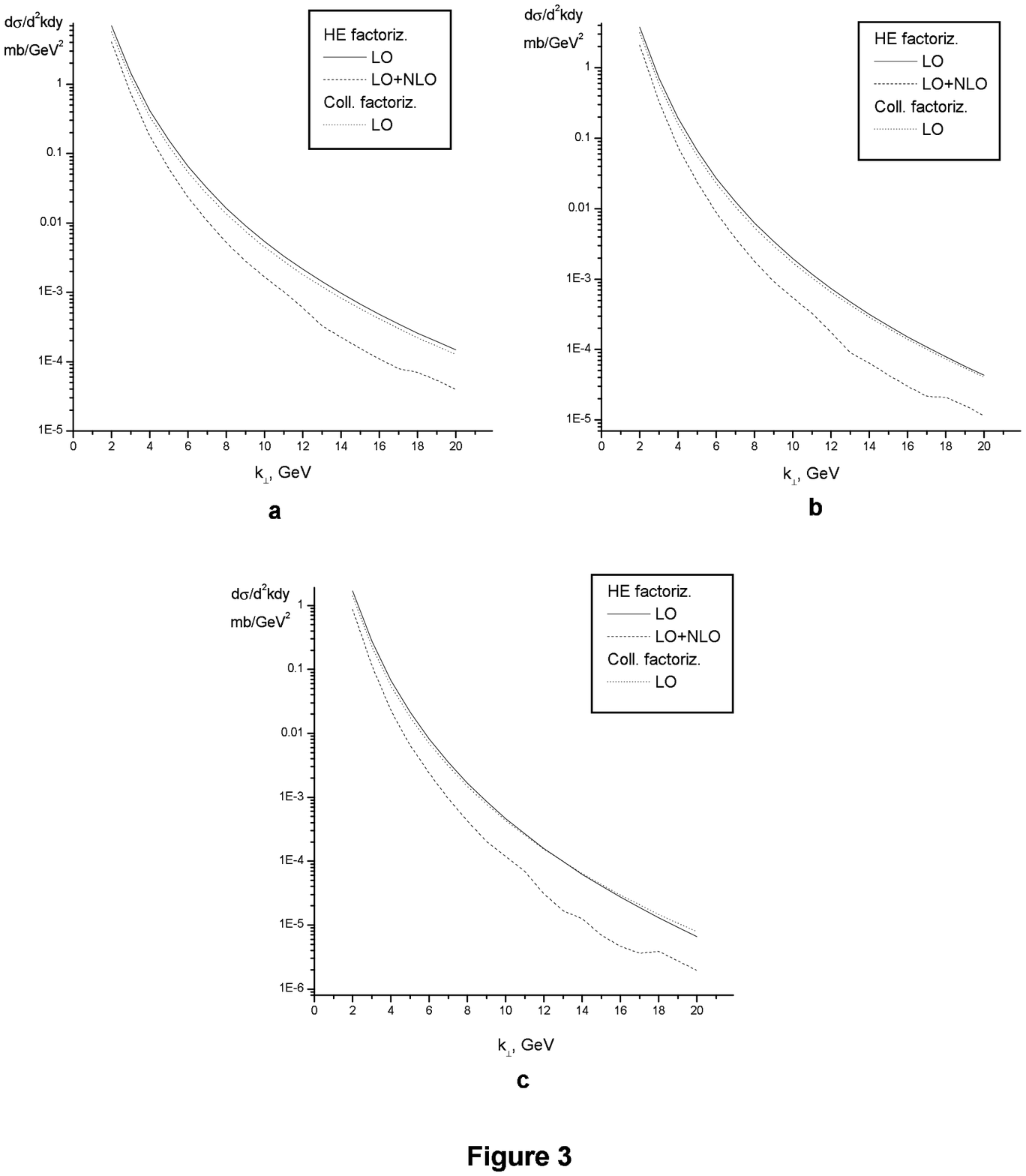,width=16cm}
\end{center}
\end{figure}

\newpage

\begin{figure}
\begin{center}
\vspace*{0cm}
\epsfig{file=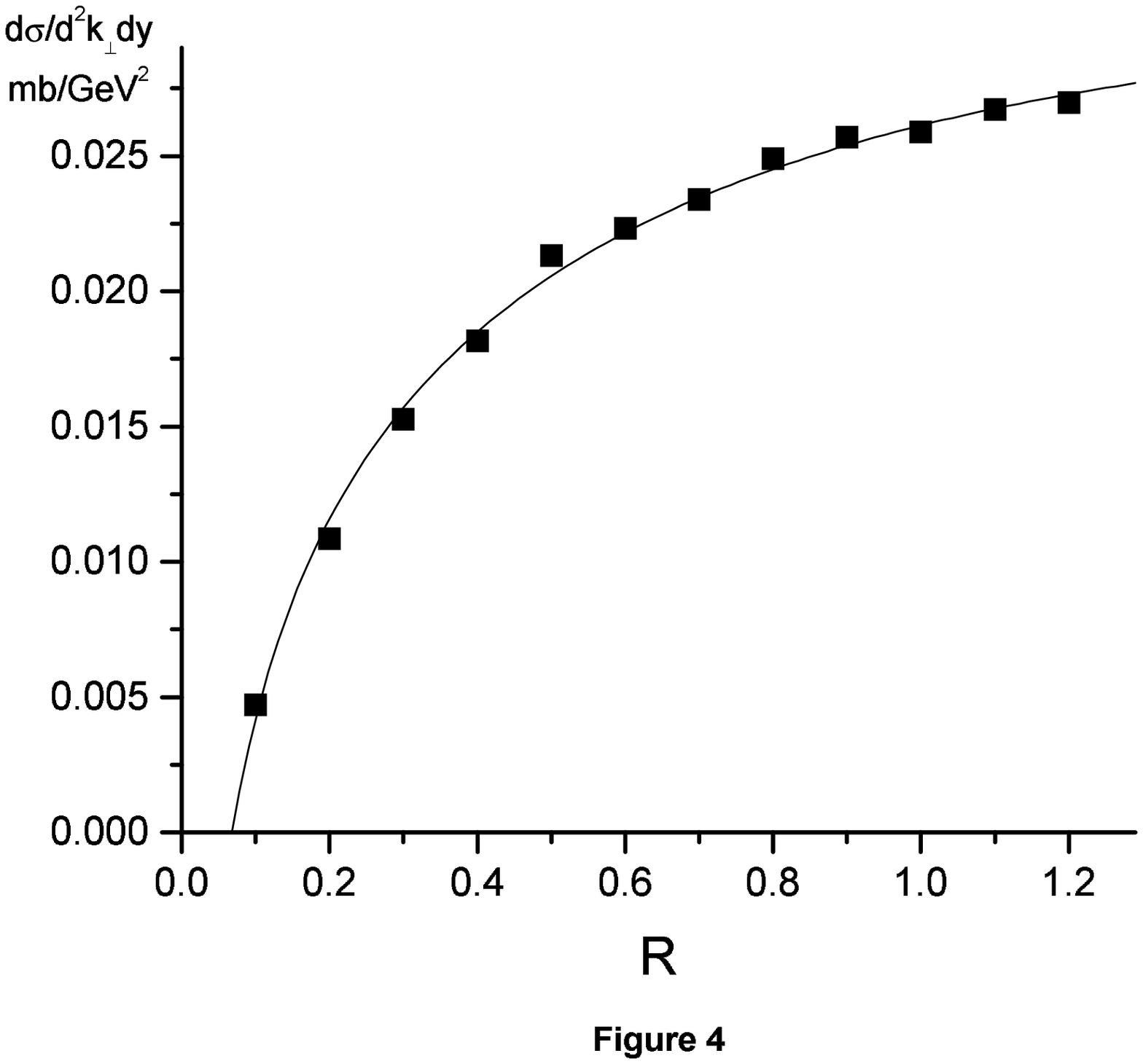,width=10cm}
\end{center}
\end{figure}

\end{document}